\documentclass[aps,prmaterials,reprint,superscriptaddress,longbibliography,floatfix]{revtex4-2}

\usepackage{amsmath}
\usepackage{amssymb}
\usepackage{graphicx}
\usepackage{booktabs}
\usepackage{multirow}
\usepackage{dcolumn}
\usepackage{xcolor}
\newcommand{\num}[1]{#1}
\newcommand{\SI}[2]{#1\,#2}
\newcommand{\SIrange}[3]{#1--#2\,#3}
\newcommand{\ang}[1]{\ensuremath{#1^\circ}}
\newcommand{\angstrom}{\AA}
\usepackage{upgreek}
\AtBeginDocument{\renewcommand{\textmu}{\ensuremath{\upmu}}}

\newcommand{\Te}{T_\mathrm{e}}
\newcommand{\mevat}{\,meV/atom}
\usepackage{etoolbox}
\makeatletter
\AtBeginDocument{\patchcmd{\NAT@thebibliography}{\clubpenalty4000\widowpenalty4000}{\clubpenalty\@M\widowpenalty\@M}{}%
  {\typeout{bib penalty patch failed}}}
\makeatother
\newsavebox{\fitbox}
\newcommand{\fitwidth}[1]{\sbox{\fitbox}{#1}%
  \ifdim\wd\fitbox>\linewidth\typeout{fitwidth: table shrunk from \the\wd\fitbox}\resizebox{\linewidth}{!}{\usebox{\fitbox}}\else\usebox{\fitbox}\fi}

\usepackage{hyperref}
\hypersetup{colorlinks=true,linkcolor=blue!60!black,citecolor=blue!60!black,urlcolor=blue!60!black,
  pdftitle={Machine-learned potentials for iron oxides at extreme conditions},
  pdfauthor={Sam Azadi}}

\begin{document}

\title{Machine-learned potentials for iron oxides at extreme conditions}
\author{S.\ Azadi}
\email{sam.azadi@manchester.ac.uk}
\affiliation{Department of Physics and Astronomy, University of Manchester, Oxford Road, Manchester M13 9PL, UK}
\affiliation{Department of Physics, Clarendon Laboratory, University of Oxford, Parks Road, Oxford OX1 3PU, UK}
\affiliation{Center for Advanced Systems Understanding, Untermarkt 20, D-02826 G\"orlitz, Germany}
\author{S.\ M.\ Vinko}
\affiliation{Department of Physics, Clarendon Laboratory, University of Oxford, Parks Road, Oxford OX1 3PU, UK}
\author{T.\ D.\ K\"{u}hne}
\affiliation{Center for Advanced Systems Understanding, Untermarkt 20, D-02826 G\"orlitz, Germany}
\affiliation{Helmholtz Zentrum Dresden-Rossendorf, Bautzner Landstra{\ss}e 400, D-01328 Dresden, Germany}
\affiliation{TU Dresden, Institute of Artificial Intelligence, Chair of Computational System Sciences, N\"othnitzer Stra{\ss}e 46 D-01187 Dresden, Germany}
\author{A.\ Principi}
\affiliation{Department of Physics and Astronomy, University of Manchester, Oxford Road, Manchester M13 9PL, UK}
\author{M.\ S.\ Bahramy}
\affiliation{Department of Physics and Astronomy, University of Manchester, Oxford Road, Manchester M13 9PL, UK}
\date{\today}

\begin{abstract}
Reliable phase stability requires accurate relative free energies, which small force errors alone do not guarantee. We develop atomic-cluster-expansion potentials for five iron oxides from Fe$_2$O to Fe$_2$O$_3$, covering crystalline and liquid states at 140--260\,GPa and 1000--11000\,K, with electronic and ionic temperatures matched. Force-dominated training leaves systematic energy offsets between structure families of up to 41\,meV/atom that persist despite additional training data. Stronger energy weighting and weaker regularization largely remove these offsets, while converged k-point sampling prevents further composition-dependent biases. The final potentials achieve held-out energy errors of 2.0--5.5\,meV/atom and first-order free-energy errors within 3.4\,meV/atom for liquids at 5000--9000\,K. However, an untrained NiAs-type FeO phase exhibits a free-energy error of 57\,meV/atom despite a low extrapolation grade. These results show that reliable phase-diagram calculations require control of relative energy offsets and explicit free-energy validation of every competing phase.
\space
\end{abstract}

\maketitle


\section{Introduction}
\label{sec:intro}
Iron oxides, Fe$_x$O$_y$, are important components of planetary materials, whose stability and physical properties influence oxygen cycling, chemical heterogeneity and transport in deep interiors. Despite their simple elemental composition, they exhibit a rich variety of stoichiometries and crystal structures under extreme conditions, accompanied by changes in coordination, spin state and electronic conductivity \cite{Grimvall,Porter,Pepperhoff,Celin26,Amouretti}. Establishing the resulting pressure--temperature--composition phase diagram therefore requires accurate free-energy differences across numerous crystalline and liquid phases, with consistent treatment of electronic excitations and magnetic properties\cite{Azadi2025II,AzadiPRM,AzadiPRR,AzadiPRB24}. Although first-principles calculations provide the underlying energetics, the system sizes and extensive statistical sampling required for converged phase boundaries make a comprehensive exploration computationally demanding. Developing accurate machine-learned interatomic potentials is therefore essential for making such calculations tractable, enabling large-scale simulations and free-energy sampling while retaining a quantitative connection to the first-principles reference \cite{cheng2019ab}.

We study iron oxides from Fe$_2$O to Fe$_2$O$_3$ at 140 to \SI{260}{GPa}, pressures of Earth's outer core
\cite{komabayashi2014thermodynamics}.
Holding 12.5\,wt\% O or more, they are richer in oxygen than estimates for the outer core \cite{hirose2021light} and
bear on oxygen-rich layers above \cite{wicks2010very} and below \cite{pozzo2019feo} the core--mantle boundary and on
the cores of super-Earths \cite{huang2018stability}, rather than on the bulk core alloy.

At high pressure the Fe--O system is richer than the FeO--Fe$_3$O$_4$--Fe$_2$O$_3$ sequence of ambient conditions,
with Fe$_4$O$_5$ \cite{lavina2011discovery}, further oxides $n$FeO$\cdot m$Fe$_2$O$_3$ above \SI{60}{GPa} and
\SI{2000}{K} and several polymorphs of Fe$_2$O$_3$ \cite{bykova2016structural}.
In earlier work we computed electronic-temperature-dependent free energies of nine crystalline phases of the five
compositions at 60 to \SI{260}{GPa} with static ions \cite{azadi2026electronic}.
Their relaxed structures are the start structures of this work, which adds the liquid of each composition.

Free energies of these phases need long simulations of large cells, beyond density-functional theory (DFT), whose
molecular dynamics (MD) of a 160-atom cell cost about 5700 core-hours per picosecond here.
Machine-learned interatomic potentials serve as intermediate Hamiltonians whose free energies are corrected to DFT
\cite{cheng2019ab,zhu2024}, which makes two demands beyond small force errors.
First, for an entropy change of one Boltzmann constant per atom, an energy error of \SI{1}{meV/atom} shifts the
temperature at which a free-energy difference vanishes by about \SI{12}{K}, consistent with estimates
\cite{zhu2024,wu2024}.
Second, the excited electrons add a large free energy that differs between phases \cite{azadi2026electronic}, so a
potential has to reproduce the Mermin free-energy surface \cite{mermin1965thermal} at an electronic temperature $\Te$
equal to the ionic temperature $T$, with $\Te$ as an input \cite{wu2024} or one potential per condition
\cite{schorner2022}.

Forces fix the energy only up to a constant \cite{chmiela2017machine}, so a force-dominated loss can leave energy
errors that force errors do not reveal, above all in training sets that mix phases \cite{kovacs2023evaluation}.
A larger force weight raised the energy error of beryllium potentials with converged labels \cite{baghishov2026},
query-by-committee selection left nearly constant energy offsets \cite{stolte2025}, and selection and stopping
criteria, an extrapolation grade \cite{lysogorskiy2023active} or committee disagreement, flag unfamiliar
configurations instead.
Labels matter too, as a $3\times3\times3$ mesh instead of the $\Gamma$ point cut the validation force error of an
aluminum potential from about 17 to 4.5\% \cite{novikov2021mlip}.
Two questions follow for datasets that mix phases and compositions, whether such a loss leaves offsets of whole
structure families while the active learning appears converged, and how far $\Gamma$-point labels in cells of
different size and composition lie from converged ones.

Machine-learned potentials with iron and oxygen exist for iron-rich liquids at outer-core conditions
\cite{zhang2022deep}, Mg--Fe--O at the core--mantle boundary \cite{deng2024} and iron oxides at ambient pressure
\cite{bienvenu2025,torres2026}, but to our knowledge none for these oxides, crystalline and liquid, at these pressures
with labels at $\Te=T$.

Here we build atomic-cluster-expansion (ACE) potentials \cite{drautz2019atomic} for these oxides at 1000 to
\SI{11000}{K}, one per electronic temperature, by active learning on DFT labels, and answer both questions.
A force-dominated loss left offsets of whole structure families, of up to \SI{41}{meV/atom} in training and held-out
structures alike, that kept the held-out energy error at 5 to \SI{17}{meV/atom} while extrapolation halts vanished,
and a strong energy weight, applied together with weaker regularization, removed most of them at force errors within
5\% of the previous ones.
At \SI{1000}{K}, k-point labels differed from $\Gamma$-point labels by $-73$ to \SI{+110}{meV/atom}, by amounts that
depend on composition and cell, and $\Gamma$-trained potentials inherited them as offsets.
The final potentials, trained on k-point labels, reach held-out energy errors of 2.0 to \SI{5.5}{meV/atom}, mostly
scatter, relative force errors of 3 to 8\% and pressure errors of at most \SI{1.4}{GPa}, and give equations of state
and, for the five liquids, pair structure and self-diffusion.

Beyond held-out structures, we test the potentials on configurations of their own MD by first-order free-energy errors
against DFT in small cells matched to the structure class of each state.
At equal ionic and electronic temperatures, these errors have a median of \SI{1.0}{meV/atom} for the
start structures that the potentials retained and stay within \SI{3.4}{meV/atom} for liquids at 5000 to
\SI{9000}{K}, whereas FeO classes not among the start structures, above all NiAs-type FeO, reach
\SI{57}{meV/atom} at extrapolation grades that do not flag them.
Free energies from these potentials are reported separately, and the Supplemental Material
\cite{SupplementalMaterial} gives the details.
\space

\section{Methods}
\label{sec:methods}

\subsection{States and DFT labels}
\label{sec:grid}
\label{sec:dft}

\begin{figure*}[tp]
\centering
\includegraphics{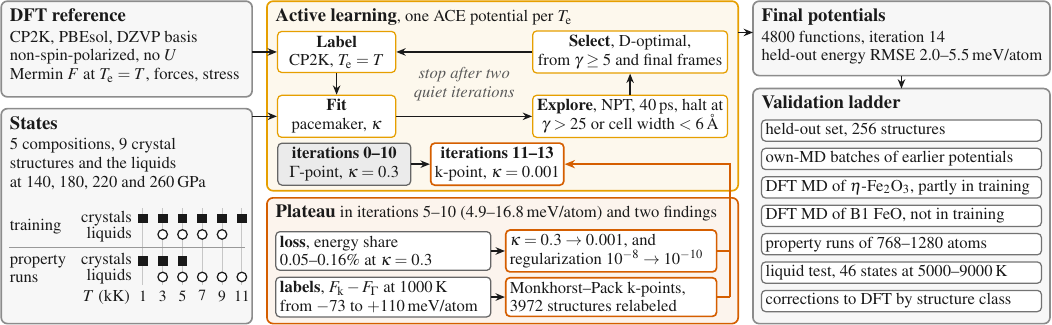}
\caption{Workflow. DFT labels with the electrons at the ionic temperature, $\Te=T$ (left, $F$ the Mermin free energy,
$U$ the Hubbard parameter), train one ACE potential per $\Te$ by active learning (center, $\gamma$ the extrapolation
grade), in stages that differ in labels, force weight $\kappa$ and basis. The map gives the temperatures of training
starts and property runs. The bottom band gives the plateau of the held-out energy error, two findings
($F_\mathrm{k}-F_\Gamma$, k-point minus $\Gamma$-point label) and their fixes, with $\kappa$ and the regularization
changed together, which lead to the k-point stage (vermilion). The final potentials (right) were tested on a ladder from
held-out structures to corrections to DFT (Secs.~\ref{sec:validation} and \ref{sec:corr}), MD meaning molecular
dynamics.}
\label{fig:workflow}
\space
\end{figure*}

We considered the nine crystalline phases of five compositions of Ref.~\cite{azadi2026electronic}, with its start
structures, namely hexagonal Fe$_2$O, B1 FeO, Fe$_4$O$_5$, Fe$_3$O$_4$ and the $\alpha$ (corundum-type), $\iota$
(Rh$_2$O$_3$(II)-type), $\eta$ (post-perovskite-type) \cite{ono2004high}, $\zeta$ (distorted perovskite) and $\theta$
(Aba2) \cite{bykova2016structural} polymorphs of Fe$_2$O$_3$, and the liquid of each composition
(Fig.~\ref{fig:workflow}).
The pressures were 140 to \SI{260}{GPa} in steps of \SI{40}{GPa} and the temperatures 1000 to \SI{11000}{K} in steps of
\SI{2000}{K}, with the electronic temperature $\Te=T$, which gives 216 crystal states and 20 composition--pressure pairs
for the liquids.
Elemental Fe and O, B2 FeO \cite{ozawa2011phase} and tetragonal Fe$_2$O \cite{jiang2026tetragonal} were not included,
and the set was not checked for completeness.

Labels were computed with CP2K 2026.2 \cite{kuhne2020cp2k}, the PBEsol functional \cite{perdew2008restoring},
Goedecker--Teter--Hutter pseudopotentials generated for PBE \cite{goedecker1996separable,krack2005pseudopotentials},
DZVP-MOLOPT basis sets \cite{vandevondele2007gaussian} and an \SI{800}{Ry} cutoff (Sec.~S1).
They were non-spin-polarized and without a Hubbard $U$, because FeO and Fe$_2$O$_3$ become metallic well below
\SI{140}{GPa}, FeO with a spin crossover \cite{ohta2012experimental} and Fe$_2$O$_3$ with a collapse of its magnetism
\cite{pasternak1999breakdown}.
With Fermi--Dirac occupations at $\Te$, each label is the Mermin free energy \cite{mermin1965thermal} with its forces
\cite{weinert1992fractional} and stress, whose term $-TS_\mathrm{el}$ reaches \SI{$-$24.5}{meV/atom} at
\SI{1000}{K} and \SI{$-$1.79}{eV/atom} at \SI{11000}{K} in a static \num{160}-atom $\eta$-Fe$_2$O$_3$ cell.
Once the $\Gamma$-point error had proved large (Sec.~\ref{sec:res-labels}), labels first computed at the $\Gamma$ point
in cells of 96 to 160 atoms were relabeled, apart from cells narrower than \SI{5.5}{\angstrom} and exact duplicates,
with Monkhorst--Pack meshes \cite{monkhorst1976special} in Minkowski-reduced cells \cite{nguyen2009low},
$3\times3\times3$ at 1000 and \SI{3000}{K} and $2\times2\times2$ above, and later structures were labeled likewise
(Sec.~S5).
At \SI{1000}{K}, $3\times3\times3$ agreed with $4\times4\times4$ to \SI{0.2}{meV/atom} in a 96-atom cell of FeO, the
composition with the largest $\Gamma$-point error, and a 112-atom Fe$_3$O$_4$ cell, and no mesh denser than
$3\times3\times3$ was tested for the training labels above \SI{1000}{K}.
Against projector-augmented-wave calculations \cite{blochl1994projector} for a 20-atom $\eta$-Fe$_2$O$_3$ cell at
\SI{140}{GPa}, the thermal free energy agreed to within \SI{4.5}{meV/atom}, but the static pressure of our protocol was
5.15 to \SI{6.13}{GPa} higher, about 0.8\% larger volumes at the same pressure for a bulk modulus of \SI{700}{GPa}.
Neither the non-magnetic choice nor this offset, measured for one composition at one pressure, was tested further, so
every error and property here, including those of B1 FeO, refers to this protocol.

\subsection{Potentials and training loss}
\label{sec:ace}

We fitted ACE potentials \cite{drautz2019atomic} with pacemaker
\cite{lysogorskiy2021performant,bochkarev2022efficient}, the model type favored by a CPU benchmark
(Sec.~S2), one potential per electronic temperature.
From active-learning iteration 5 on, the basis had 2400 functions (\num{11124} parameters, cutoff
\SI{5.5}{\angstrom}) instead of 844, and 4800-function bases extended 2400-function potentials by new functions
starting at zero.
The loss is
\begin{equation}
  \mathcal{L} = (1-\kappa)\sum_i w^E_i \left(\frac{\Delta E_i}{N_i}\right)^2 + \kappa \sum_{i,a} w^F_{ia}\,|\Delta\mathbf{F}_{ia}|^2
  + \mathcal{L}_\mathrm{reg},
  \label{eq:loss}
\end{equation}
with structure and atom weights $w^E$ and $w^F$ from the energy-based weighting policy of pacemaker (energy weights
spanning a factor of 9 to 17 in the iteration-8 data), each set normalized to one, so that $\kappa$ is the relative
weight of the forces, and with L1, L2 and radial-smoothness terms in $\mathcal{L}_\mathrm{reg}$.
The potentials trained on $\Gamma$-point labels (iterations 0 to 10) used $\kappa=0.3$ and regularization coefficients
of $10^{-8}$, the refits of the iteration-8 data $\kappa=0.003$ or 0.001 and $10^{-10}$, and those trained on k-point
labels (iterations 11 to 14) $\kappa=0.001$ and $10^{-10}$.
BFGS in SciPy \cite{virtanen2020scipy} ran for 1000 steps in fits from scratch and after a change of labels or basis,
and for 400 otherwise (Sec.~S3).
Stress was not fitted, and the pressure was constrained by volume-varied training structures instead, which in the
benchmark reduced the pressure bias of an energy-and-force fit from $-23$ to \SI{$+$0.25}{GPa}.
Per atom and step, the 4800-function potentials cost 347 to \SI{378}{core\text{-}\textmu s}, 1.5 to
$2.1\times10^{5}$ times less than DFT molecular dynamics (Sec.~S3).

\subsection{Training data and active learning}
\label{sec:al}

\looseness=-1 DFT molecular dynamics (DFT-MD) of a \num{160}-atom $\eta$-Fe$_2$O$_3$ cell at \SI{140}{GPa}, \SI{2}{ps} each at
1000, 6000 and \SI{11000}{K} at the $\Gamma$ point, supplied training frames and thermal expansions
(Sec.~S1).
At 6000 and \SI{11000}{K} both species diffused and the cell lost its orthorhombic shape.
For every phase, pressure and temperature, three copies of a cell of 96 to 160 atoms were expanded thermally, scaled by
up to $\pm1.5\%$, strained (standard deviation 0.004) and rattled by 0.06 to \SI{0.26}{\angstrom} from 1000 to
\SI{11000}{K}, giving 648 seed structures.
Copies 0 and 1 and 203 DFT-MD frames at 1000 and \SI{11000}{K}, 60 of them volume-scaled, were training data, and copy
2 was held out.

Each iteration ran \SI{40}{ps} of NPT molecular dynamics in LAMMPS \cite{thompson2022lammps} with a Nos\'e--Hoover
thermostat and barostat \cite{nose1984molecular,hoover1985canonical,martyna1994constant} from every crystal start and,
from iteration 3 on, from liquid starts.
Frames with a largest extrapolation grade $\gamma\geq5$ \cite{lysogorskiy2023active,podryabinkin2017active} and the
final frames were candidates, and a run stopped at $\gamma>25$.
Both thresholds are heuristic.
From iteration 5 on, runs also stopped at a cell width below \SI{6}{\angstrom}, and liquid and, later, hot crystal
starts, which had drifted into thin slabs, used an isotropic barostat (Sec.~S3).
Up to 40 candidates per temperature (80 at 5000 to \SI{9000}{K} from iteration 2 on) were selected by MaxVol
D-optimal selection \cite{goreinov2010how,lysogorskiy2023active}, Niggli-reduced \cite{grossekunstleve2004numerically}
and labeled.
An iteration was quiet if no temperature had more than two extrapolation halts and every held-out energy error changed
by less than \SI{0.5}{meV/atom}, and two quiet iterations in a row stopped the loop.
Iterations 11 to 13, with k-point labels, 2400 functions and $\kappa=0.001$, also required that no run was stopped by
a time or memory limit, and the loop stopped after the quiet iterations 12 and 13 (Table~S3).
Because the extrapolation grade depends on the basis, these data were then refitted with 4800 functions and checked by
one exploration, which was not quiet, so one more selection, labeling and 400-step refit gave the final potentials of
iteration 14, explored once more as a check (Sec.~\ref{sec:res-final}).

\subsection{Validation sets, error measures and baselines}
\label{sec:validation}

Copy 2 of every seed state and 40 DFT-MD frames, 16 at \SI{1000}{K} and 24 at \SI{11000}{K}, formed a fixed held-out
set of 256 structures, which also served the stop rule and the choice of $\kappa$ and basis.
Its seed structures are near-twins of training copies, and it holds no liquid apart from the \SI{11000}{K} frames and
no configuration of the potentials' own molecular dynamics.
Such configurations were tested by the structures selected for labeling, before they entered training, 332 from the
explorations of the $\kappa=0.3$ potentials of iteration 8, 315 and 323 from iterations 11 and 12, and 340 from the
check exploration.
These own-MD batches are own-MD data only for the potentials that explored them.
The final potentials were tested on own-MD configurations by the corrections of Sec.~\ref{sec:corr} and, at
$T=\Te=5000$, 7000 and \SI{9000}{K}, by a liquid test of 46 states of the five liquids at 140 to \SI{260}{GPa} in the
protocol of the liquid corrections.
Of these states, 24 are the liquid corrections at 5000 and \SI{7000}{K} of Sec.~\ref{sec:res-corr}.

Every error in this paper is the potential minus DFT, $\Delta X=X_\mathrm{pot}-X_\mathrm{DFT}$, including the
first-order free-energy error of Sec.~\ref{sec:corr}.
The energy error is the root-mean-square error (RMSE) of $\Delta E/N$, offsets included, force RMSEs are per Cartesian
component with structures weighted equally (relative to the root-mean-square DFT component where stated), and
pressures are $-\mathrm{tr}\,\sigma/3$ of the static stress $\sigma$.
The 12 seed structures of one phase at one temperature form a family, and the mean $\Delta E/N$ of its training or
held-out members is a family offset.
For $M$ structures with errors $e_i$ in families $g$ of $n_g$ members with mean errors $\mu_g$,
\begin{equation}
  \begin{aligned}
  \mathrm{RMSE}^2 &= O^2 + S^2, \\
  O^2 &= \frac{1}{M}\sum_g n_g\,\mu_g^2, \quad
  S^2 = \frac{1}{M}\sum_g\sum_{i\in g}\left(e_i-\mu_g\right)^2 ,
  \end{aligned}
  \label{eq:split}
\end{equation}
splits the RMSE exactly into the root-mean-square family offset $O$ and the scatter $S$ within families.
Scatter alone gives offsets of about $s/\sqrt{n_g}$, with $s$ the pooled standard deviation within families
(noise-corrected offset $O_\mathrm{c}$, Sec.~S6).
Equation~(\ref{eq:split}) was applied per temperature to the 36 held-out seed structures (nine families of four), and to
the own-MD batches with compositions as groups.
Changes between potentials were judged by paired bootstraps over whole families, or over structures within
compositions for the own-MD batches, and constant-energy null models set the scale (Sec.~S6).
No committee or ensemble uncertainty was computed.
\space

\subsection{Corrections to DFT and structure classes}
\label{sec:corr}
\label{sec:target}

Free-energy perturbation \cite{zwanzig1954high} gives the free-energy error of a potential in a state,
\begin{equation}
\frac{F_\mathrm{pot}-F_\mathrm{DFT}}{N}=\frac{k_\mathrm{B}T}{N}\ln\bigl\langle e^{\delta/k_\mathrm{B}T}\bigr\rangle_\mathrm{pot}
\ge\epsilon_1\equiv\frac{\langle\delta\rangle_\mathrm{pot}}{N},
\label{eq:fep}
\end{equation}
where $\delta=U_\mathrm{pot}-U_\mathrm{DFT}$ is the energy of the potential minus the DFT energy (the Mermin free energy
at $\Te$) of the same configuration of $N$ atoms, and $\langle\cdot\rangle_\mathrm{pot}$ averages over molecular dynamics
with the potential at the ionic temperature $T$.
By the Gibbs--Bogoliubov inequality \cite{isihara1968gibbs}, the first-order error $\epsilon_1$ bounds the signed
free-energy error per atom from below, so a negative $\epsilon_1$ can overstate its magnitude, and the second-order
(cumulant) estimate adds $\sigma_\delta^2/2k_\mathrm{B}TN$, with $\sigma_\delta$ the standard deviation of $\delta$ in
the cell.
Both need $\sigma_\delta$ of order $k_\mathrm{B}T$ or less, which per atom of a 160-atom cell is \SI{0.54}{meV} at
\SI{1000}{K} and \SI{1.6}{meV} at \SI{3000}{K}.

The states, from free-energy calculations with the final potentials reported separately, lie at 1000 to 2125,
1000 to 3625, 3000 to 5000 and \SI{7000}{K} for the potentials of 1000, 3000, 5000 and \SI{7000}{K}.
In the 188 of the 304 states with $T\neq\Te$, $\epsilon_1$ still compares with DFT at the $\Te$ of the potential and
so tests its transfer to configurations hotter or colder than those of its active learning.
It leaves out the error of describing the electrons at $\Te$ instead of $T$.
For each crystal state a cell of 96 to 320 atoms was folded from the production cell (Sec.~S10), and each
liquid used a small melt at its mean volume.
After \SI{10}{ps} of NVT dynamics with the potential of the state, six frames \SI{1}{ps} apart (more in seven states
and five in one, Sec.~S10) received CP2K single points in the label protocol (Sec.~\ref{sec:dft}).
These fixed-volume errors would change only at second order in the pressure error at fixed pressure.

Structure classes were assigned by one rule module per composition from the space group (spglib
\cite{togo2024spglib}) and the Steinhardt bond-order parameters $q_4$ and $q_6$ \cite{steinhardt1983bond} of
time-averaged positions (Sec.~S10).
In FeO, structures with equivalent Fe and O sublattices form the rocksalt family, with the classes B1, its monoclinic or
orthorhombic distortion mB1, and rB1 and rB1$'$, elongated and compressed along [111].
A crystal correction was kept only if its frame-averaged small cell had the class, and the symmetry variant within the
class, of its state (the frame audit, Sec.~S10).
We grouped the states into start classes, the classes of the active-learning start structures (Sec.~\ref{sec:al}), other
classes and liquids.
\space
\section{Results}
\label{sec:results}


\subsection{A force-dominated loss leaves per-family energy offsets}
\label{sec:res-offsets}
\label{sec:res-al}
\label{sec:res-kappa}

\begin{figure*}[tp]
\centering
\includegraphics{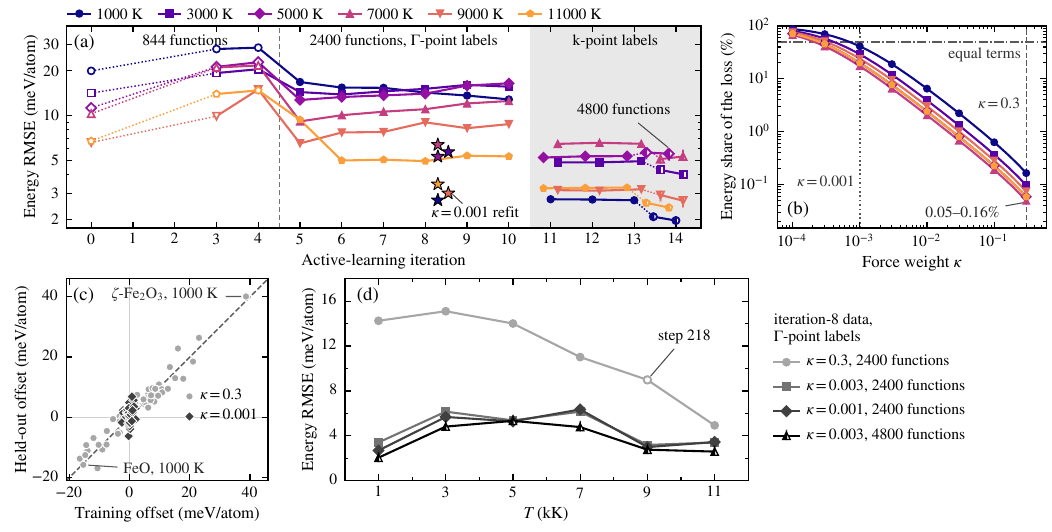}
\caption{Energy weight and family offsets, one potential per temperature $T$ ($\Te=T$).
(a)~Held-out energy RMSE over the iterations (not evaluated at 1 and 2), against $\Gamma$-point labels to iteration 10
and k-point labels from 11 on (shaded). Open markers show 844 functions, filled ones 2400 and half-filled ones 4800 (the
iteration-13 refit and the final potentials), stars the $\kappa=0.001$ refit of iteration 8.
(b)~Energy share of the loss of Eq.~(\ref{eq:loss}) against $\kappa$, computed at the errors of the $\kappa=0.3$ fits
held fixed.
(c)~Offset of each seed family on its held-out against its training structures, $y=x$ dashed.
(d)~Held-out energy RMSE of the iteration-8 potentials and their refits, the open circle a fit stopped at step 218.}
\label{fig:weight}
\label{fig:stall}
\space
\end{figure*}

Over iterations 0 to 10 the extrapolation halts fell from 177 runs to none (Table~S3), a sign of
coverage, not of energy accuracy.
From iteration 5 to 10, with 2400 functions, the held-out energy error stayed at 4.9 to \SI{16.8}{meV/atom} while the
training set grew by 70\% to \num{3525} structures [Fig.~\ref{fig:weight}(a)], and the stop rule did not fire, because
from iteration 7 on the error still moved by up to \SI{1.9}{meV/atom} between iterations.

With $\kappa=0.3$, nominally an energy weight of 0.7, the force residuals dominated the loss numerically
[Fig.~\ref{fig:weight}(b)].
The energies made up 0.05 to 0.16\% of the final loss of the iteration-8 fits, and the two terms would have been equal
only at $\kappa\approx2$--$7\times10^{-4}$.
Longer fits would have gained little, as the training energy error changed by $-0.09$ to
\SI{$-$0.22}{meV/atom} over their last 100 steps.
Most of the held-out error was a constant offset shared by all configurations of one family, including its training
configurations [Fig.~\ref{fig:weight}(c)], such as \SI{$+$40.1}{meV/atom} on the held-out and \SI{$+$38.7}{meV/atom}
on the training structures of $\zeta$-Fe$_2$O$_3$ at \SI{1000}{K}.
Forces constrain energy differences only along sampled paths, so they cannot detect offsets between families that the
training data do not connect \cite{chmiela2017machine}, and the extrapolation grade, low by construction on training
configurations, did not flag them either.
By Eq.~(\ref{eq:split}) the offset $O$ of the iteration-8 potentials, 4.6 to \SI{16.0}{meV/atom}, exceeded the scatter
$S$, 3.5 to \SI{8.2}{meV/atom}, at every temperature (Table~S8).

We therefore refitted the iteration-8 potentials on the same data in the same basis with $\kappa=0.003$ and 0.001 (400
steps), lowering at the same time the regularization from $10^{-8}$ to $10^{-10}$, whose L1 and L2 terms had been 5 to
33\% of the energy term, so that the weaker regularization may share in the result, a confound not removed.
With $\kappa=0.001$ the held-out energy error fell from 4.9--15.1 to 2.7--\SI{6.4}{meV/atom}
[Fig.~\ref{fig:weight}(d)], and the relative force errors stayed within 5\% of their previous values.
The offset $O$ fell to 1.2 to \SI{3.2}{meV/atom}, at most \SI{2.1}{meV/atom} after the noise correction, whereas $S$ fell
by more than \SI{1}{meV/atom} only at 1000, 3000 and \SI{9000}{K}, so the refit removed mostly the per-family offsets.
The family-cluster bootstrap resolves the fall at all six temperatures, with seven to nine of nine families lower
(Table~S8).
On the 332 structures selected from the explorations of the $\kappa=0.3$ potentials, the refit lowered the error from
15.5 to \SI{4.3}{meV/atom} at \SI{1000}{K} and from 10.1 to \SI{6.6}{meV/atom} at
\SI{3000}{K}, resolved also at 7000 and \SI{9000}{K}, while at \SI{11000}{K} it rose from 5.6 to \SI{5.8}{meV/atom}
(Tables~S8 and S9).
Doubling the basis at $\kappa=0.003$ lowered the held-out energy error by at most \SI{1.4}{meV/atom}, against 1.5 to
\SI{11.6}{meV/atom} from the change of the loss, with basis size and fit length not separated (Sec.~S4).
\space

\subsection{\texorpdfstring{$\Gamma$}{Gamma}-point labels carry composition- and cell-dependent biases}
\label{sec:res-labels}
\label{sec:res-diag}
\label{sec:res-kpt}

\begin{figure}[tbp]
\centering
\includegraphics{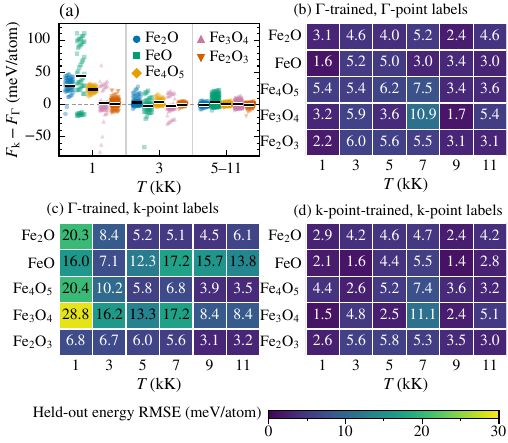}
\caption{$\Gamma$-point against k-point labels.
(a)~k-point minus $\Gamma$-point label, $F_\mathrm{k}-F_\Gamma$ (Mermin free energy per atom), of the \num{3962}
structures with both labels at 1000, 3000 and 5000 to \SI{11000}{K} (meshes $3\times3\times3$ at 1000 and
\SI{3000}{K}, $2\times2\times2$ above), by composition, with composition means as bars.
(b)--(d)~Held-out energy RMSE by composition and temperature $T$ of 2400-function potentials with $\kappa=0.001$,
trained on $\Gamma$-point labels and tested on $\Gamma$-point (b) or k-point labels (c), and trained and tested on
k-point labels (d), from four structures per composition and temperature (20 to 44 for Fe$_2$O$_3$).}
\label{fig:labels}
\space
\end{figure}

For the \num{3962} structures with both labels, the \num{3972} relabeled ones except ten without a $\Gamma$-point label
[Fig.~\ref{fig:labels}(a), Table~S6], the k-point label differs from the $\Gamma$-point label at
\SI{1000}{K} by $-72.7$ to \SI{$+$110.0}{meV/atom} (standard deviation 21.9,
537 structures), with composition means from \SI{$+$0.9}{meV/atom} for Fe$_2$O$_3$ to \SI{$+$43.9}{meV/atom} for FeO
and a standard deviation of \SI{16.2}{meV/atom} within compositions.
Each composition except Fe$_2$O$_3$ used one cell size, and the 120- and 160-atom Fe$_2$O$_3$ cells have means of
$-8.3$ and \SI{$+$1.6}{meV/atom}, so the dependence on composition is not separated from one on the cell.
The standard deviation is \SI{7.4}{meV/atom} at \SI{3000}{K} and 2.3 to \SI{2.9}{meV/atom} at 5000 to \SI{11000}{K}.
It is a bias of the labels, not a shown cause of the plateau of Sec.~\ref{sec:res-offsets}, since in checked cells at
\SI{1000}{K} it explained a fifth to a half of the variance of the model error within compositions, while at 5000 and
\SI{11000}{K} the model error did not follow it (Sec.~S5).
It would shift free-energy differences between compositions by up to \SI{44}{meV/atom} (FeO against Fe$_2$O$_3$).

The $\kappa=0.001$ potentials trained on $\Gamma$-point labels are 2.7 to \SI{6.4}{meV/atom} from these labels but 5.3
to \SI{13.4}{meV/atom} from the k-point labels of the held-out structures.
At \SI{1000}{K} the error is \SI{6.8}{meV/atom} for Fe$_2$O$_3$ but 16 to \SI{29}{meV/atom} for the other compositions
[Fig.~\ref{fig:labels}(b), (c)].
On k-point labels their offset is \SI{15.3}{meV/atom} at \SI{1000}{K} and 5.1 to \SI{7.7}{meV/atom} at the other
temperatures, against 1.2 to \SI{3.2}{meV/atom} on their own labels, while $S$ changes by less than
\SI{1}{meV/atom} [Fig.~\ref{fig:final}(b)], so the bias of the labels enters the potentials as family offsets.

Potentials fitted to the k-point labels (2400 functions, $\kappa=0.001$, 1000 steps from the $\Gamma$-trained ones) are
2.7 to \SI{6.4}{meV/atom} from them, without the composition dependence [Fig.~\ref{fig:labels}(d)], and the
family-cluster bootstrap resolves the fall except at 7000 and \SI{9000}{K}.
These fits also had 853 more training structures and 1000 more steps, a confound not removed, but the
$\Gamma$-trained potentials reach this level on their own labels without them, so the extra data and steps are not
needed for it.
On the 315 structures selected from the explorations of the k-point-trained potentials, the $\Gamma$-trained potentials
err at \SI{1000}{K} by \SI{26.6}{meV/atom} and the k-point-trained ones by \SI{8.4}{meV/atom}.
FeO and Fe$_2$O are off by $-46.9$ and \SI{$-$33.4}{meV/atom} with the $\Gamma$-trained potentials, close to minus their
mean $F_\mathrm{k}-F_\Gamma$, and this fall, with the same confound, is resolved only at 1000 and \SI{11000}{K}
(Table~S8).
\space

\subsection{Final potentials and their errors on unseen configurations}
\label{sec:res-final}
\label{sec:res-kal}
\label{sec:res-xl}

\begin{figure*}[tp]
\centering
\includegraphics{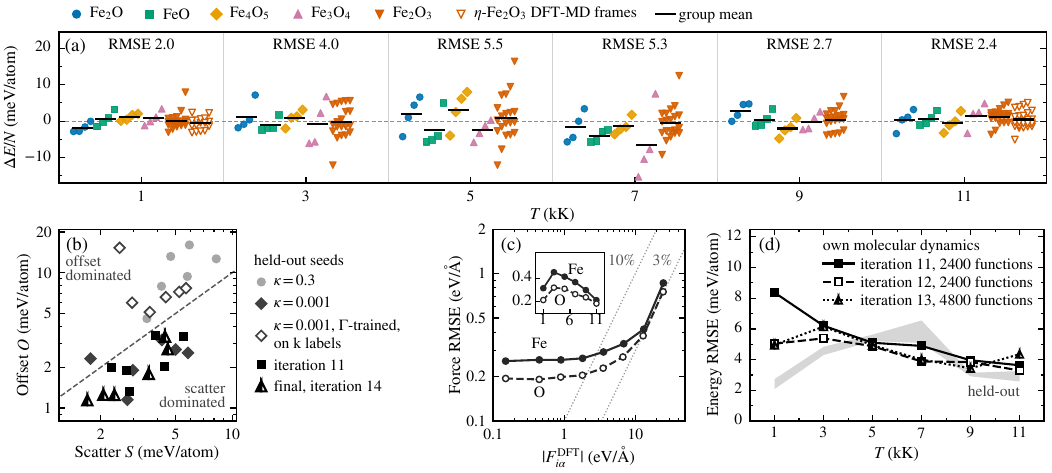}
\caption{The final potentials and errors on unseen configurations.
(a)~Energy error $\Delta E/N$ of the final potentials on each held-out structure by composition and temperature $T$,
with group means as bars and the RMSE of each $T$ in meV/atom.
(b)~Offset $O$ against scatter $S$ of Eq.~(\ref{eq:split}) on the held-out seeds, one point per $T$, for the
potentials of Table~S8, with $O=S$ dashed.
(c)~Force RMSE of Fe and O atoms in bins of $|F^\mathrm{DFT}_{i\alpha}|$ over all $T$, dotted at 3 and 10\% of
$|F^\mathrm{DFT}_{i\alpha}|$, and against $T$ (inset).
(d)~Energy RMSE on configurations of the own molecular dynamics of earlier potentials, not the final ones, against the
held-out RMSE of the same potentials (band).}
\label{fig:final}
\space
\end{figure*}

\begin{table}[tb]
\caption{Root-mean-square errors of the final potentials, potential minus DFT, at $T=\Te$, as defined in
Sec.~\ref{sec:validation}. The held-out rows cover the 256 held-out structures, with the shear stress over the three
off-diagonal components. The own-MD rows, in the units of the held-out rows, cover frames of the potentials' own
molecular dynamics from the corrections of Sec.~\ref{sec:res-corr} at 1000 and \SI{3000}{K} (22 and 17 start-class
states, 20 liquid states) and from the liquid test at 5000, 7000 and \SI{9000}{K} (18, 18 and 10 states, forces pooled
over atoms). Section~\ref{sec:res-corr} gives the errors of other classes.}
\label{tab:final}
\fitwidth{%
\begin{tabular}{l*{6}{c}}
\toprule
$T$ (kK) & 1 & 3 & 5 & 7 & 9 & 11 \\
\midrule
\multicolumn{7}{l}{\emph{held-out structures}} \\
structures & 52 & 36 & 36 & 36 & 36 & 60 \\
energy (meV/atom) & 2.0 & 4.0 & 5.5 & 5.3 & 2.7 & 2.4 \\
force (meV/\AA) & 264 & 381 & 359 & 317 & 267 & 198 \\
relative force (\%) & 7.9 & 5.4 & 4.4 & 3.9 & 3.2 & 3.0 \\
pressure (GPa) & 1.15 & 1.30 & 1.16 & 0.97 & 0.85 & 0.54 \\
shear stress (GPa) & 0.60 & 0.49 & 0.50 & 0.47 & 0.36 & 0.25 \\
\multicolumn{7}{l}{\emph{own molecular dynamics at $T=\Te$}} \\
energy, start classes & 1.5 & 3.2 & -- & -- & -- & -- \\
energy, liquids & -- & 4.1 & 4.4 & 3.7 & 3.2 & -- \\
force, liquids & -- & -- & 301 & 273 & 236 & -- \\
pressure, liquids & -- & -- & 0.79 & 0.57 & 0.40 & -- \\
\bottomrule
\end{tabular}
}
\end{table}

In the active learning with k-point labels, the held-out energy errors of the iteration-13 potentials lie within
\SI{0.11}{meV/atom} of those of iteration 11 (Table~S7).

The structures selected in each iteration test the potentials that explored them before they enter training
[Fig.~\ref{fig:final}(d), Table~S9], and as they were chosen by extrapolation grade and D-optimality, they
lean toward configurations unfamiliar to these potentials.
The 2400-function potentials of iterations 11 and 12 erred by 3.6 to 8.4 and 3.3 to \SI{5.4}{meV/atom} on the 315 and
323 structures selected from their explorations, against 2.7 to \SI{6.5}{meV/atom} on the held-out set, most above
it at \SI{1000}{K} (8.4 and 5.0 against \SI{2.7}{meV/atom}), and from \SI{3000}{K} on their composition offsets
were at most \SI{3.0}{meV/atom} (Table~S8).
The structures selected from the check exploration of the 4800-function refit entered the training set of the final
potentials, so no own-MD batch tests these.

The final potentials [Table~\ref{tab:final}, Fig.~\ref{fig:final}(a)--(c)] have held-out energy errors of 2.0 to
\SI{5.5}{meV/atom}.
At 5000 to \SI{11000}{K} the held-out structures are crystalline seeds, and at \SI{11000}{K} also frames of the DFT
trajectory that supplied training frames.
On frames of their own molecular dynamics at $T=\Te$, from the corrections of Sec.~\ref{sec:res-corr} at 1000 and
\SI{3000}{K} and the liquid test at 5000 to \SI{9000}{K}, the energy errors of start classes and liquids are 1.5 to
\SI{4.4}{meV/atom}, at most \SI{0.5}{meV/atom} above the held-out error at the same temperature, and the
\SI{11000}{K} potential has no such test.
The held-out force errors are 3.0 to 7.9\% of the root-mean-square DFT force, larger for Fe (214 to
\SI{450}{meV/\angstrom}) than for O (185 to \SI{320}{meV/\angstrom}), also at the same DFT force, but relative to the
force scale of each species they differ by at most 12\%.
The held-out pressure errors are at most \SI{1.4}{GPa}, and those of the off-diagonal stress components at most
\SI{0.7}{GPa}.
With four structures per composition and temperature (20 to 44 for Fe$_2$O$_3$), the errors by composition are 1.3 to
\SI{10.7}{meV/atom}, and the largest held-out family offset is \SI{$+$7.6}{meV/atom} ($\zeta$-Fe$_2$O$_3$,
\SI{5000}{K}).
The offset $O$ is 1.1 to \SI{3.4}{meV/atom}, or 0 to \SI{2.2}{meV/atom} after the noise correction, and the scatter
1.7 to \SI{4.5}{meV/atom}, so the scatter dominates at every temperature [Fig.~\ref{fig:final}(b)].
\space

\subsection{Equations of state, structure and transport}
\label{sec:res-prop}

\begin{figure*}[tb]
\centering
\includegraphics{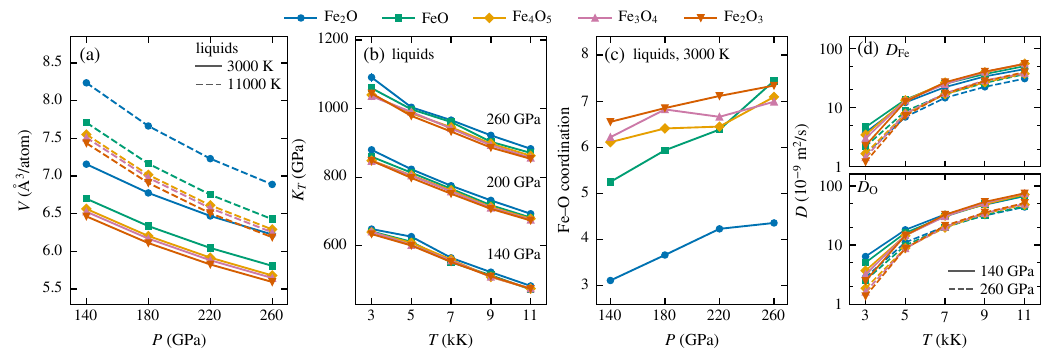}

\caption{Equations of state, structure and transport of the liquids with the final potentials at $\Te=T$.
(a) Volume per atom at 3000 and \SI{11000}{K}.
(b) Isothermal bulk modulus at 140, 200 and \SI{260}{GPa} from the Vinet form referenced at \SI{200}{GPa}, with error
bars of the statistical error scaled by the square root of $\chi^2$ per degree of freedom (at most \SI{14}{GPa}, mostly
within the symbols).
(c) Fe--O coordination at \SI{3000}{K}.
(d) Self-diffusion coefficients of Fe and O at 140 and \SI{260}{GPa} from multiple time origins, block errors
within the symbols, compared by composition in Table~S11.
Values in Tables~S10 and S11.}
\label{fig:prop}
\space
\end{figure*}

The property runs used the final potentials at $\Te=T$ in $2\times2\times2$ replicas of the active-learning start cells
(768 to \num{1280} atoms) with \SI{10}{ps} of equilibration and \SI{40}{ps} of production at constant pressure, at 100
liquid and 108 crystal states and 36 crystal restarts at \SI{1000}{K} (Sec.~S7).
All liquid runs and 73 of the 140 finished crystal runs passed their mobility gates (crystals immobile, liquids
diffusive), and whether a liquid is the equilibrium phase at its state is not addressed.
Crystal properties come only from the 31 runs at 1000 and \SI{3000}{K} that kept their start structure and passed the
gate, at extrapolation grades of at most 2.14 (Table~S10).
The largest grades of the liquid runs, maxima over all atoms, were 1.25 to 16.0, above the candidate threshold of 5 in
33 of 100 runs, none at \SI{3000}{K} (Table~S11).

The isothermal bulk modulus $K_T$ of the liquids at \SI{200}{GPa}, from the Vinet form \cite{vinet1987temperature}
referenced there, falls from 847--880 at \SI{3000}{K} to \SIrange{674}{694}{GPa} at \SI{11000}{K}, monotonically for
every composition (Fig.~\ref{fig:prop}).
This form and the third-order Birch--Murnaghan form \cite{birch1947finite} miss some volumes by up to four block
errors \cite{flyvbjerg1989error}, so the statistical errors are scaled by the square root of $\chi^2$ per degree of
freedom, to at most 4, 9 and \SI{14}{GPa} at 200, 140 and \SI{260}{GPa}.
The seven crystal series with three or four pressures give $K_T$ of 902 to \SI{1029}{GPa} at \SI{200}{GPa}, with
scaled errors of at most \SI{9}{GPa} (Table~S10), and neither form fits $\eta$- and $\iota$-Fe$_2$O$_3$
at \SI{1000}{K} (Sec.~S7), so their errors understate the uncertainty.

At \SI{3000}{K} the Fe--O coordination of the liquids rises with the O content, from 3.11 (Fe$_2$O) to 6.55
(Fe$_2$O$_3$) at \SI{140}{GPa}, and with pressure, by 0.77 (Fe$_3$O$_4$) to 2.19 (FeO) up to \SI{260}{GPa},
monotonically except for Fe$_3$O$_4$ at \SI{220}{GPa} (Table~S11).
At higher temperatures it is not interpreted, as the first minimum of $g_\mathrm{FeO}$ is flat.
At every temperature the first Fe--O peak, at 1.71 to \SI{1.79}{\AA}, moves from 140 to \SI{260}{GPa} by at most two bins
(\SI{0.054}{\AA}), less than a uniform compression would (0.08 to \SI{0.10}{\AA}).

Self-diffusion coefficients from multiple time origins, with block errors of 0.2 to 6\%, range from 1.21 to 55.7 for
Fe and from 1.36 to 75.2 for O, in $10^{-9}$\,m$^2$/s.
They fall with pressure and rise with temperature in every liquid and carry no finite-size correction
\cite{yeh2004system}, which is 5 to 7\% for 1280 atoms at \SI{10816}{K} (Sec.~S7).
O diffuses faster than Fe in every state, with $D_\mathrm{O}/D_\mathrm{Fe}={}$1.39 to 1.80 in Fe$_2$O and 1.05 to 1.36
in the other liquids.

All values are those of the potentials, on the pressure scale of the CP2K protocol (Sec.~\ref{sec:dft}).
The five liquids are nearly alike, expanding from 3000 to \SI{11000}{K} by 15.0 to 15.1\% at \SI{140}{GPa} and 10.6
to 10.8\% at \SI{260}{GPa} with $K_T$ within 5.4\% of each other, which is not interpreted, as stress was not fitted.
In the liquid test of Sec.~\ref{sec:res-final}, the first-order errors were within \SI{3.4}{meV/atom}, the cumulant
estimates within \SI{7.5}{meV/atom} (median 2.2), and the pressure errors $-1.2$ to \SI{+1.0}{GPa} per state at
grades of at most 3.2, whereas the property runs of these states reached
16.0 (Sec.~S7).
\space

\subsection{The potentials against DFT molecular dynamics}
\label{sec:res-val}

The DFT molecular dynamics of the 160-atom $\eta$-Fe$_2$O$_3$ cell at \SI{140}{GPa} (Sec.~\ref{sec:al}), at the
$\Gamma$ point, supplied 203 training and 40 held-out frames, so they test whether the potentials reproduce trajectories
they were partly trained on, including the diffusive state at \SI{10816}{K}.
On these frames the k-point labels differ from the $\Gamma$-point ones by $+3.9$ and \SI{0.0}{meV/atom} and
$+0.17$ and \SI{+0.002}{GPa} (985 and \SI{10816}{K}, Sec.~S8).
Runs (i) and (iii) of the final potentials at the same electronic temperatures started from the first DFT frame, in the
same cell and its 1280-atom replica.
\begin{table}[tb]
\caption{The potentials against DFT molecular dynamics at \SI{140}{GPa}, as potential minus DFT or potential/DFT with
standard errors (Sec.~S8).
Upper part, $\eta$-Fe$_2$O$_3$, runs (i) and (iii) against the 160-atom $\Gamma$-point DFT runs, which supplied
training frames, with mean-square displacements (MSD) about the mean positions at \SI{985}{K} and self-diffusion
coefficients $D$ at \SI{10816}{K}.
Lower part, the 96-atom FeO cell started as B1, bond orders of the time-averaged positions and MSD from the start, for
run (a), the median of it and ten replicas, and the cell-size run (without bond orders).}
\label{tab:dftmd}
\setlength{\tabcolsep}{2.5pt}
\fitwidth{%
\begin{tabular*}{\columnwidth}{@{\extracolsep{\fill}}l*{4}{D{.}{.}{1.6}}}
\toprule
$\eta$-Fe$_2$O$_3$ & \multicolumn{2}{c}{\SI{985}{K}} & \multicolumn{2}{c}{\SI{10816}{K}} \\
\cmidrule(lr){2-3}\cmidrule(lr){4-5}
run & \multicolumn{1}{c}{(i)} & \multicolumn{1}{c}{(iii)} & \multicolumn{1}{c}{(i)} & \multicolumn{1}{c}{(iii)} \\
\midrule
$\Delta V/V$ (\%) & +0.19(12) & +0.09(11) & +0.01(99) & +0.06(82) \\
rms $\Delta g$, Fe--Fe & 0.073 & 0.119 & 0.058 & 0.049 \\
rms $\Delta g$, Fe--O & 0.059 & 0.066 & 0.043 & 0.034 \\
rms $\Delta g$, O--O & 0.044 & 0.066 & 0.041 & 0.037 \\
$\Delta n_\mathrm{FeO}$ & -0.04(2) & -0.03(1) & -0.06(9) & -0.01(6) \\
MSD or $D$, Fe & 0.97(11) & \multicolumn{1}{c}{--} & 0.92(14) & \multicolumn{1}{c}{--} \\
MSD or $D$, O & 0.90(7) & \multicolumn{1}{c}{--} & 1.04(11) & \multicolumn{1}{c}{--} \\
\bottomrule
\end{tabular*}%
}

\vspace{6pt}
\fitwidth{%
\begin{tabular*}{\columnwidth}{@{\extracolsep{\fill}}l*{4}{D{.}{.}{2.4}}}
\toprule
FeO, \SI{1000}{K} & \multicolumn{1}{c}{DFT} & \multicolumn{1}{c}{(a)} & \multicolumn{1}{c}{11 runs} & \multicolumn{1}{c}{size run} \\
averaged (ps) & \multicolumn{1}{c}{1--2} & \multicolumn{1}{c}{1--2} & \multicolumn{1}{c}{1--2} & \multicolumn{1}{c}{3--100} \\
\midrule
$q_4$ of Fe & 0.1985 & 0.2073 & 0.2200 & \multicolumn{1}{c}{--} \\
$q_4$ of O & 0.1977 & 0.2081 & 0.2200 & \multicolumn{1}{c}{--} \\
$q_6$ of Fe & 0.4896 & 0.5196 & 0.4672 & \multicolumn{1}{c}{--} \\
$q_6$ of O & 0.4565 & 0.5180 & 0.4606 & \multicolumn{1}{c}{--} \\
Fe with $<6$ O (\%) & 10.4 & 0 & 0 & 0 \\
MSD of Fe, 1\,ps (\AA$^2$) & 0.21 & 0.16 & 0.15 & 0.14 \\
MSD of O, 1\,ps (\AA$^2$) & 0.31 & 0.16 & 0.16 & 0.18 \\
MSD of Fe, 2\,ps (\AA$^2$) & 0.21 & 0.26 & 0.15 & 0.60 \\
MSD of O, 2\,ps (\AA$^2$) & 0.26 & 0.28 & 0.16 & 0.88 \\
\bottomrule
\end{tabular*}%
}
\end{table}

The volumes differ by at most 0.20\%, or 1.6 standard errors of the larger estimate (Sec.~S8), the
partial $g(r)$ below
\SI{4.0}{\AA} differ by root-mean-square values of 0.034 to 0.119, and the Fe--O coordination at the first minimum of
the DFT $g_\mathrm{FeO}$ differs by at most 0.07 (Table~\ref{tab:dftmd}, Fig.~S4).
The Fe--O peak lies in the DFT bin at \SI{985}{K} and three bins lower at \SI{10816}{K}.
Run (i) has $0.97\pm0.11$ (Fe) and $0.90\pm0.07$ (O) times the DFT mean-square displacements about the mean positions
at \SI{985}{K}, with block errors that are lower bounds, and $0.92\pm0.14$ and $1.04\pm0.11$ times the DFT
self-diffusion coefficients at \SI{10816}{K}.
The grade along these runs reached at most 1.8 at \SI{985}{K} and 6.1 at \SI{10816}{K}.
Run (ii), \SI{20}{ps} in the 160-atom cell at \SI{985}{K}, moved its atoms by \SI{1.9}{\AA$^2$} and one cell angle
to \ang{102}, while run (iii) kept its atoms near their start sites, a finite-size observation of one trajectory,
beyond the \SI{2}{ps} of DFT.
Configurations outside the training data are tested against DFT in Sec.~\ref{sec:res-feo}, apart from its B1 seeds,
by the liquid test of Sec.~\ref{sec:res-final} and in Sec.~\ref{sec:res-corr}.
\space

\subsection{B1 FeO in larger cells}
\label{sec:res-feo}

In the property runs the energy collapsed, with atoms lost, only at \SI{1000}{K}, in the 768-atom B1 FeO cell at
\SI{140}{GPa} and in three restarts, B1 FeO at 140 and \SI{180}{GPa} and $\eta$-Fe$_2$O$_3$ at \SI{260}{GPa}.
Evaluated afterward on the frames of the failing B1 run, the grade was already 43 to 54, above the halt threshold of
25 (Sec.~\ref{sec:al}), when production began at \SI{10}{ps}, at least \SI{18}{ps} before the energy collapsed at
about \SI{28}{ps}, and it exceeded $10^3$ from \SI{16}{ps}.
In \SI{100}{ps} cell-size runs at \SI{140}{GPa}, B1 cells of 96, 216 and 512 atoms kept grades of at most 1.9 at 1000
and \SI{3000}{K}, whereas in the 768-atom cell at \SI{1000}{K} the grade exceeded 25 after \SI{5}{ps}, also with the
iteration-13 potentials, for a reason not established (Sec.~S9).

In the non-spin-polarized PBEsol protocol at $\Te={}$\SI{1000}{K}, the potentials and DFT agree that B1 is not a static
minimum.
Trigonal strains of 4\% in $c/a$ grew at 140 and \SI{260}{GPa}, and at \SI{140}{GPa} a 64-atom cell strained by up to
1.5\% and rattled by up to \SI{0.08}{\AA} relaxed away from B1 in DFT, whereas ideal 8-atom B1 cells kept their
cubic cell only by symmetry (Sec.~S9).
Experiments find B1 FeO up to \SI{156}{GPa} and \SI{3100}{K} \cite{fischer2011equation}, so this result concerns the
static surface of the protocol.

DFT molecular dynamics of the 96-atom cell at \SI{1000}{K}, from the start of the cell-size run (Sec.~S8),
keeps two equivalent sublattices for \SI{2.1}{ps}, but its positions averaged over 1 to \SI{2}{ps} have a mean $q_6$
of the two sublattices of 0.47, below the B1 window of the class rule (0.49) and the 0.57 of the start sites, and
10.4\% of its Fe atoms have fewer than six O neighbors within \SI{2.3}{\AA}, against none in the runs of the potential
(Table~\ref{tab:dftmd}).
Its successive \SI{0.5}{ps} averages lie 0.32 to \SI{0.36}{\AA} apart (Sec.~S8).
Run (a) of the potential and ten more runs with new velocities from this start move their atoms less than in DFT at
\SI{1}{ps} and 0.4 to 1.3 times as far at \SI{2}{ps}, and none leaves the rocksalt family of the rule within
\SI{3}{ps} (95\% upper bound of 0.28 on the fraction within \SI{2.1}{ps}), whereas the cell-size run left it by
\SI{2}{ps}.
They differ from DFT in the cell, one angle of which exceeds the largest deviation from \ang{90} of any DFT frame
(\ang{5.9}) within 0.5 to \SI{1.6}{ps} in ten of the 11 runs and reaches about \ang{13}, a shear that one DFT
trajectory of \SI{2.1}{ps} can neither confirm nor exclude.

On FeO seed configurations at 140 to \SI{260}{GPa}, the potential minus DFT is $-5.6$ to $+4.9$\mevat{} for B1 but
$-60.2$ to $+9.5$\mevat{} for B8 (NiAs type) and $-7.1$ to $+19.2$\mevat{} for trigonally strained B1, both relaxed
with the potentials (Sec.~S9).
On cells relaxed in DFT it is $-1.2$ and $+1.7$\mevat{} for B1 but $+51$\mevat{} for the rattled 64-atom B1 cell
and $+108$ for B8,
whose $c/a$ of 2.21 differs from the 1.98 of the potential's own B8, and $-48$ and $-90$\mevat{} for B2, with pressure
errors of $-28$ to $+20$\,GPa.
Their grades are 0.65 to 1.51, like those of the start structures (Sec.~\ref{sec:res-prop}), except 4.45 and 4.85
for B2.
Errors of tens of meV/atom outside the start set thus occur at familiar grades (Sec.~\ref{sec:res-corr}), whereas in
the larger B1 cell the grade rose long before the collapse and has to be monitored in production runs.
\space

\subsection{First-order free-energy errors of the potentials by structure class}
\label{sec:res-corr}

\begin{figure*}[tbp]
\centering
\includegraphics{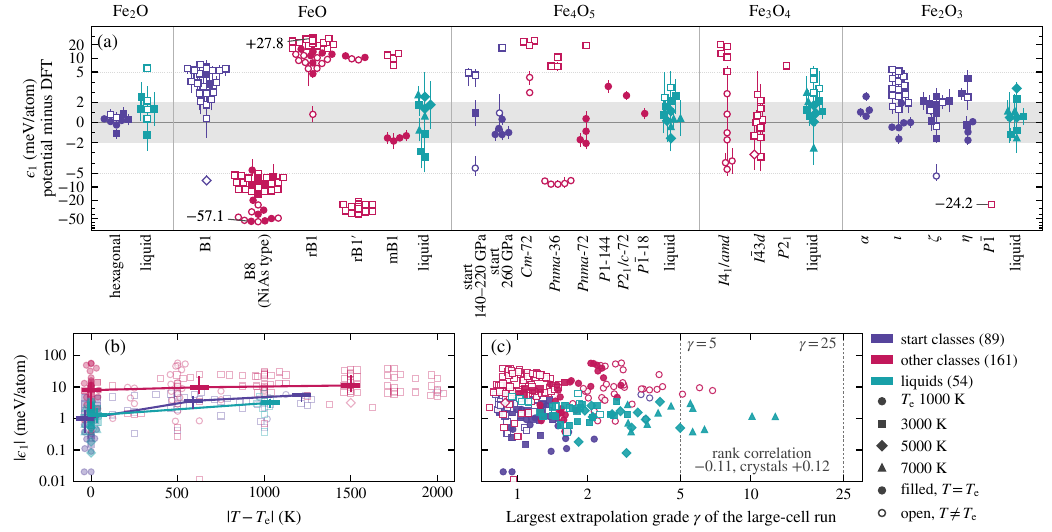}
\caption{First-order free-energy errors $\epsilon_1$ of the final potentials, potential minus DFT
[Eq.~(\ref{eq:fep})], in 304 states.
(a)~$\epsilon_1$ with its standard error, on an axis linear within $\pm$\SI{5}{meV/atom} and logarithmic beyond,
with a $\pm$\SI{2}{meV/atom} band, about \SI{23}{K} for an entropy change of $k_\mathrm{B}$ per atom.
Fe$_4$O$_5$ labels end with the atom count per conventional cell.
(b)~$|\epsilon_1|$ against $|T-\Te|$, with group medians (horizontal bars, joined) and quartiles (vertical bars) at
$|T-\Te|=0$, at most \SI{1000}{K} and beyond.
(c)~$|\epsilon_1|$ against the largest grade $\gamma$ of the large-cell run (the labeled frames were not graded).}
\label{fig:corr}
\space
\end{figure*}

\begin{table}[tb]
\caption{First-order errors $\epsilon_1$ of the final potentials, potential minus DFT, in meV/atom.
Here $n$ counts the states, $|\epsilon_1|$ and $|\epsilon_\mathrm{cum}|$ are the median absolute first-order and
cumulant errors, and at $T=\Te$ the median standard error of $\epsilon_1$ is 0.6, 0.7 and 1.5 for start classes, other
classes and liquids.}
\label{tab:corr}
\setlength{\tabcolsep}{3.2pt}%
\fitwidth{%
\begin{tabular}{lrrrrrc}
\toprule
 & \multicolumn{3}{c}{$T=\Te$} & \multicolumn{2}{c}{$T\neq\Te$} & all $T$ \\
\cmidrule(lr){2-4}\cmidrule(lr){5-6}\cmidrule(lr){7-7}
 & $n$ & $|\epsilon_1|$ & $|\epsilon_\mathrm{cum}|$ & $n$ & $|\epsilon_1|$ & range of $\epsilon_1$ \\
\midrule
\multicolumn{7}{l}{Fe$_2$O} \\
\hspace{0.6em}start classes & 7 & 0.3 & 0.6 & 2 & 0.7 & $-1.1$ to $+0.9$ \\
\hspace{0.6em}liquid & 4 & 1.4 & 4.8 & 4 & 1.7 & $-1.2$ to $+6.1$ \\
\multicolumn{7}{l}{FeO} \\
\hspace{0.6em}start classes & 2 & 4.3 & 6.2 & 24 & 5.1 & $-7.2$ to $+7.3$ \\
\hspace{0.6em}other classes & 26 & 10.9 & 12.1 & 89 & 12.3 & $-58.2$ to $+27.8$ \\
\hspace{0.6em}liquid & 10 & 1.8 & 2.6 & 0 & -- & $-3.4$ to $+2.8$ \\
\multicolumn{7}{l}{Fe$_4$O$_5$} \\
\hspace{0.6em}start classes & 6 & 1.0 & 0.8 & 5 & 4.7 & $-4.5$ to $+17.0$ \\
\hspace{0.6em}other classes & 7 & 1.7 & 3.3 & 14 & 8.6 & $-8.7$ to $+24.3$ \\
\hspace{0.6em}liquid & 10 & 0.9 & 2.6 & 2 & 3.0 & $-1.6$ to $+3.5$ \\
\multicolumn{7}{l}{Fe$_3$O$_4$} \\
\hspace{0.6em}other classes & 0 & -- & -- & 24 & 2.8 & $-4.6$ to $+20.3$ \\
\hspace{0.6em}liquid & 10 & 1.5 & 2.1 & 4 & 3.8 & $-2.5$ to $+5.0$ \\
\multicolumn{7}{l}{Fe$_2$O$_3$} \\
\hspace{0.6em}start classes & 24 & 1.4 & 1.8 & 19 & 2.8 & $-5.8$ to $+5.7$ \\
\hspace{0.6em}other classes & 0 & -- & -- & 1 & 24.2 & $-24.2$ \\
\hspace{0.6em}liquid & 10 & 1.2 & 4.0 & 0 & -- & $-1.5$ to $+3.4$ \\
\multicolumn{7}{l}{\emph{all compositions}} \\
\hspace{0.6em}start classes & 39 & 1.0 & 1.5 & 50 & 3.9 & $-7.2$ to $+17.0$ \\
\hspace{0.6em}other classes & 33 & 8.0 & 5.5 & 128 & 10.3 & $-58.2$ to $+27.8$ \\
\hspace{0.6em}liquids & 44 & 1.3 & 2.8 & 10 & 3.2 & $-3.4$ to $+6.1$ \\
\multicolumn{7}{l}{\emph{FeO excluded}} \\
\hspace{0.6em}start classes & 37 & 1.0 & 1.2 & 26 & 3.1 & $-5.8$ to $+17.0$ \\
\hspace{0.6em}other classes & 7 & 1.7 & 3.3 & 39 & 4.6 & $-24.2$ to $+24.3$ \\
\bottomrule
\end{tabular}
}
\end{table}
\providecommand{\CorrDataVersion}{907ca9f3c0282dd1b8b9dd2c9cc4a21e}
\space

The corrections of Sec.~\ref{sec:corr} test the final potentials on configurations of their own molecular dynamics,
outside the training set, in 304 states, 250 of crystals and 54 of liquids, at 140 to \SI{260}{GPa} and ionic
temperatures of \num{1000} to \SI{7000}{K} (Fig.~\ref{fig:corr} and Table~\ref{tab:corr}).
Their first-order errors depend on the class and grow as the ionic temperature departs from the $T=\Te$ of active
learning.

Where $T=\Te$, the median $|\epsilon_1|$ is \SI{1.0}{meV/atom} for the 39 start-class states, 21 of them within two
standard errors of zero and all within \SI{4.8}{meV/atom}, \SI{1.3}{meV/atom} for the 44 liquids of the 3000 to
\SI{7000}{K} potentials and \SI{1.7}{meV/atom} for the seven Fe$_4$O$_5$ states of other classes, but
\SI{10.9}{meV/atom} for the 26 FeO states of other classes, the largest being \SI{$-57.1$}{meV/atom} for B8 (NiAs
type) at \SI{140}{GPa}.
These start-class values cover only start structures kept by the dynamics of the potentials, with no state of the
Fe$_3$O$_4$ start structure, one of $\theta$-Fe$_2$O$_3$ that failed the frame audit and none of B1 with the
\SI{1000}{K} potential.
At $T=\Te$ the cumulant estimate has median magnitudes of 1.5, 5.5 and \SI{2.8}{meV/atom} for start classes, other
classes and liquids, so the contrast stays, while the start-class and liquid errors are known only to a few meV/atom,
since six frames fix $\sigma_\delta^2$ only to about $\pm60\%$.
Without the 124 states with $\sigma_\delta/k_\mathrm{B}T$ above 2 [Fig.~S5(b)], the medians at
$T=\Te$ are 0.9 (29 start-class states) and \SI{10.1}{meV/atom} (12 other-class states).
Pooled medians weigh classes by their number of states in the free-energy calculations (FeO holds 115 of the 161
other-class states), but over classes the medians at $T=\Te$ are 1.1 and \SI{3.1}{meV/atom} for start and other
classes (Table~S12).

Away from $\Te$, the medians are \SI{3.9}{meV/atom} for the start classes (50 states) and \SI{10.3}{meV/atom} for
the other classes (128), and for the start classes the median grows from 1.0 at $T=\Te$ to 3.6 and \SI{5.5}{meV/atom}
at $|T-\Te|$ of at most \SI{1000}{K} and beyond [Fig.~\ref{fig:corr}(b)].
Of the 82 states above \SI{10}{meV/atom}, 81 belong to other classes, 68 of them to FeO B8, rB1 and rB1$'$, while 34
other-class states lie within \SI{5}{meV/atom}.
Two start classes have larger errors, FeO B1, offset at every $|T-\Te|$ by medians of \SI{$+4.3$}{meV/atom} at $T=\Te$
(two states) and \SI{$+4.4$}{meV/atom} otherwise (24), and the Fe$_4$O$_5$ start class of \SI{260}{GPa}, which
reaches \SI{$+17.0$}{meV/atom} at \SI{260}{GPa} and \SI{1500}{K} with the \SI{3000}{K} potential.
FeO B8 errs by a median of \SI{$-43.1$}{meV/atom} with the \SI{1000}{K} potential (14 states) and
\SI{$-7.6$}{meV/atom} with the \SI{3000}{K} potential (37), so a correction belongs to one potential.

Per atom, $\sigma_\delta$ has a median of \SI{2.0}{meV/atom} in start and other classes alike, also when scaled by
$N^{1/2}$ (24.9 and \SI{24.1}{meV}), comparable in size to the held-out scatter of 1.7 and \SI{3.6}{meV/atom} at 1000
and \SI{3000}{K} (Table~S8), so the other classes are offset rather than noisier, as the family offsets
of Sec.~\ref{sec:res-offsets}.
The pressure, recorded for 144 states (Sec.~S10), differs from DFT by medians of $-0.1$, $-0.3$ and
\SI{$-0.2$}{GPa} for start classes, other classes and liquids, at most \SI{8.5}{GPa} or about 1\% in volume,
against held-out pressure errors of up to \SI{1.4}{GPa} (Sec.~\ref{sec:res-final}).
The 82 states above \SI{10}{meV/atom} reached extrapolation grades of at most 4.2 in the large-cell runs of their
states, below the candidate threshold of 5, and grade and $|\epsilon_1|$ are nearly uncorrelated [rank correlation
$-0.11$, crystals $+0.12$, Fig.~\ref{fig:corr}(c)].
The grade thus finds their local environments close to the training data, whose structures were not classified,
while the energy of the whole structure is off.
Denser k-point meshes changed $\epsilon_1$ by at most \SI{0.22}{meV/atom} in six single points on one frame each of
four cells of 112 to 224 atoms (Fe$_3$O$_4$ $P2_1$ and $I4_1/amd$ and two of FeO mB1) and were not tested on the B8,
rB1, rB1$'$ and 288-atom Fe$_4$O$_5$ cells.
Corrections of the same class and potential in different cell sizes at neighboring states, within \SI{20}{GPa} and
\SI{250}{K} of each other, differ by at most 3.0 and \SI{4.0}{meV/atom} for $\iota$-Fe$_2$O$_3$ (160 and 320 atoms)
and Fe$_4$O$_5$ \mbox{$Pnma$-36} (144 and 288 atoms), the only classes with such pairs.
\space
\space

\section{Conclusions and recommendations}
\label{sec:conclusions}

A force-dominated loss left family offsets, invisible to force errors, that kept the held-out energy error at 4.9 to
\SI{16.8}{meV/atom} while extrapolation halts vanished, and a refit with a much larger energy weight and weaker
regularization removed most of them (Sec.~\ref{sec:res-offsets}).
At \SI{1000}{K}, k-point labels differed from $\Gamma$-point labels by $-73$ to \SI{+110}{meV/atom}, by
composition- and cell-dependent amounts that $\Gamma$-trained potentials inherited as offsets
(Sec.~\ref{sec:res-labels}).
The final potentials reach held-out energy errors of 2.0 to \SI{5.5}{meV/atom}, mostly scatter, and give equations of
state, Fe--O coordination and self-diffusion of the liquids (Secs.~\ref{sec:res-final} and \ref{sec:res-prop}).
They reproduce DFT molecular dynamics of $\eta$-Fe$_2$O$_3$, whose trajectories supplied training frames, in volume
within 0.20\%, at most 1.6 standard errors of the larger estimate, in Fe--O coordination within 0.07 and in
mean-square displacement and self-diffusion within 10\% (Sec.~\ref{sec:res-val}, Table~\ref{tab:dftmd}).
First-order free-energy errors against DFT on configurations of their own molecular dynamics depend on the structure
class and grow as the ionic temperature departs from the $T=\Te$ of active learning (Sec.~\ref{sec:res-corr}).
At $T=\Te$, the first-order errors of the start classes retained by the potentials have a median of
1.0\,meV/atom, about 1.5 at second order, and those of liquids at 5000 to \SI{9000}{K} stay within
\SI{3.4}{meV/atom}, whereas FeO classes outside the start set reach 57\,meV/atom at extrapolation grades
below the candidate threshold.
On cells relaxed in DFT the potentials err by $-48$ and \SI{$-90$}{meV/atom} for B2 FeO, outside the start set.
In the non-spin-polarized PBEsol protocol, B1 FeO is not a static minimum for the potentials or DFT.
In a 96-atom B1 cell, whose DFT molecular dynamics is independent of the training data, eleven runs of the potential
from the DFT start keep two equivalent sublattices like DFT, but their time-averaged structures have no Fe atom with
fewer than six O neighbors, against 10\% in DFT, and ten shear the cell more (Sec.~\ref{sec:res-feo},
Table~\ref{tab:dftmd}).
In a 768-atom B1 cell the extrapolation grade, evaluated afterward on the saved frames, exceeded the halt threshold at
least \SI{18}{ps} before the energy collapsed.

For free-energy work we recommend energy-weighted losses \cite{kovacs2023evaluation}, but only with converged labels
\cite{baghishov2026}, k-point sampling converged for every cell, energy errors split into offsets and scatter with
family intervals, and free-energy validation of every structure class used with the potential of its electronic
temperature.
Seeds should be relaxed with broken symmetry, since in DFT relaxations B1 FeO kept its cubic cell only by symmetry
(Secs.~\ref{sec:res-feo} and S9), and the extrapolation grade should be monitored in production runs,
above all in hot liquids and cells larger than the training cells.
Free energies from these potentials are reported separately.
\space

\begin{acknowledgments}
S.A, A.P, and M.S.B acknowledge the support of the Leverhulme Trust under the grant agreement RPG-2023-253. S. Azadi and T.D. K\"{u}hne acknowledge the computing time provided to them on the high-performance computers Noctua2 at the NHR Center in Paderborn (PC2). S.M.V. acknowledges support from the UK EPSRC under grant EP/W010097/1.
\end{acknowledgments}

\section*{Data availability}
The data are not publicly available. The data are available from the authors upon reasonable request.


\bibliography{main}
\space

\end{document}